\documentclass[twocolumn,showpacs,preprintnumbers,amsmath,amssymb]{revtex4}
\usepackage{graphicx}
\usepackage{dcolumn}
\usepackage{bm}

\newcommand{\be}{\begin{equation}}
\newcommand{\ee}{\end{equation}}
\newcommand{\ba}{\begin{eqnarray}}
\newcommand{\ea}{\end{eqnarray}}

\begin{document}
\renewcommand{\figurename}{{\bf Fig.}}
\renewcommand{\tablename}{{\bf Tab.}}
\renewcommand{\figurename}{{\bf Fig.}}
\renewcommand{\tablename}{{\bf Tab.}}

\title{Quantum entanglement dynamics and decoherence wave in spin
chains at finite temperatures}
\author{S. D. ~Hamieh and M. I. ~Katsnelson}
\affiliation{Institute for Molecules and Materials, Radboud
University of Nijmegen, 6525 ED Nijmegen, The Netherlands}
\date{\today}
\begin{abstract}
We analyze the quantum entanglement at the equilibrium in a class
of exactly solvable one-dimensional spin models at finite
temperatures and identify a region where the quantum fluctuations
determine the behavior of the system. We probe the response of the
system in this region by studying the spin dynamics after
projective measurement of one local spin which leads to the
appearance of the ``decoherence wave''. We investigate
time-dependent spin correlation functions, the entanglement
dynamics, and the fidelity of the quantum information transfer
after the measurement.
\end{abstract}
\pacs{03.65.Ud; 03.67.Mn; 03.65.Ta}
\maketitle

\section{Introduction}
\label{sect:1}

Collective behavior in many-body quantum systems is associated
with the development of classical correlations, as well as of the
correlations which cannot be accounted for in terms of classical
physics, namely, entanglement. The entanglement represents in
essence the impossibility of giving a local description of a
many-body quantum state. Experimental tests of the nonlocality by
means of the Bell-type inequality \cite{1} have been made with
different kind of particles including photons \cite{2} and massive
fermions \cite{3,4,44}. The entanglement is expected to play an
essential role at quantum phase transitions \cite{5}, where
quantum fluctuations manifest themselves at all length scales.
Several groups investigated this problem by studying the quantum
spin systems (see, e.g.,
Refs.\onlinecite{Osbo02,19,Hami05,12,14,24,25,26,27,28,29,30,31,32,Amic04,Subr05,ghos1}).
Additional studies have been carried out for more complicated
systems including both itinerant electrons and localized spins;
the local entanglement for these systems have been discussed in a
context of the quantum phase transitions \cite{gu,Anfo} and of the
Kondo problem \cite{ourkondo}. In particular, Anfossi {\it et al}
\cite{Anfo} performed, within the density-matrix renormalization
group method, a numerical comparison between the standard
finite-size scaling and the local entanglement for the Hubbard
model in a presence of bond charge interaction (the Hirsch model)
at the Mott metal-insulator transition. Katsnelson {\it et al}
\cite{ourkondo} have considered the suppression of the Kondo
resonance by a probing of the charge state of the magnetic
impurity which leads to the partial destruction of the
entanglement between the localized spin and itinerant-electron
Fermi sea. These examples illustrate a relevance of the concept of
entanglement for the many-body physics. Moreover, the entanglement
overwhelmingly comes into play in the quantum computation and
communication theory \cite{6}, being the main physical resource
needed for their specific tasks. The essential idea is to encode
one particular qubit, and let it be transported to across the
chain to recover the code from another qubit some distance away
\cite{7}.

The suppression of the entanglement by decohering actions such as
noise, measurements, etc., is one of the central problems in
quantum computation and quantum information theory; therefore the
concept of entanglement for mixed states is of primary relevance
\cite{Benn96}. For example, it is important to know what happens
with the quantum computer after the measurement of one qubit
state; for the case of the quantum system with broken continuous
symmetry such as Bose-Einstein condensate (BEC) or easy-plane
antiferromagnet the local measurements lead to the formation of
the ``decoherence wave'' \cite{ourbec,ourneel}. It is interesting
to investigate the effect of the decoherence wave on the
entanglement in the system.

Motivated by these results, in this paper, we aim on the
evaluation of the pairwise entanglement in the 1D Ising-XY model
with transverse magnetic field at finite temperatures. We identify
a region where the thermal entanglement is non zero while it is
zero at zero temperature, which results from the entanglement of
the excited states as it will be explained below
(section~\ref{sect:2}). We study the dynamical response of the
system in the non vanishing entanglement region, which is the
useful region for quantum information processing, after a
projective measurement on one local spin. We find that, similar to
the case of the BEC considered earlier \cite{ourbec} the spin
decoherence wave appears propagating with the velocity
proportional to the interaction strength. We investigate also the
zero temperature case studying the time-dependent correlation
functions and we discuss the relation between the dynamics of the
magnetization and the entanglement (section~\ref{sect:3}). Our
conclusions are given in section~\ref{sect:4}.

%%%%%%%%%%%%%%%%%%%%%%%%%%%%%%%%%%%%%%%%%%%%%%%%%%%%%%%%%%%%%%%%%%%%%%%%%%%%%%%%%%%%%%%%%%%%%%%%%%%%%%%
\section{Thermal entanglement in the Ising-XY Model with transverse field}
\label{sect:2}

\begin{figure}[htb]
\centering\includegraphics[width=9cm]{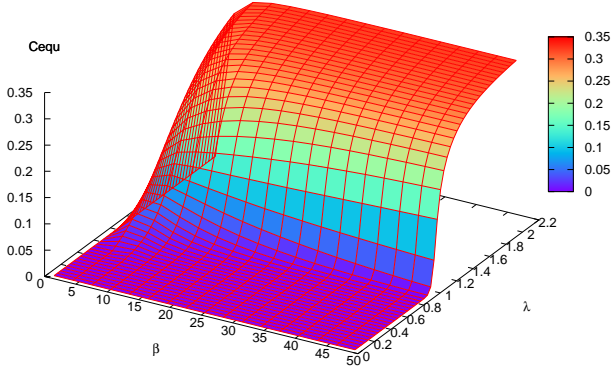}\vspace{0cm}
\caption{(color online) Pairwise entanglement for the nearest neighbors in the 
isotropic XY model with transverse field at the equilibrium as function of
$\beta$ and $\lambda$. Cequ is defined by Eq.(12) for the thermodynamic
equilibrium state.\protect\label{Cequ1}}
\end{figure}

\begin{figure}[htb]
 \centering\includegraphics[width=9cm]{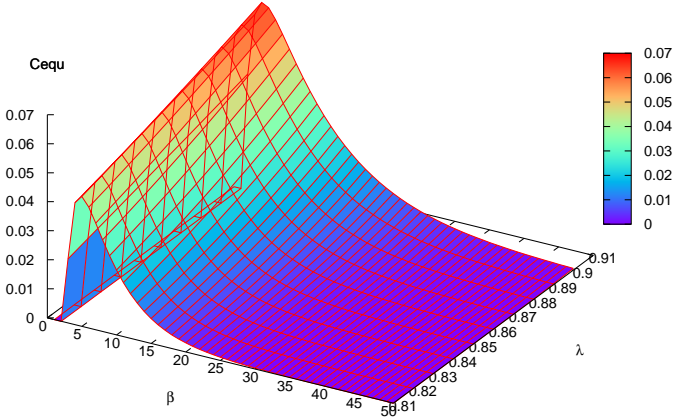}\vspace{0cm}
\caption{(color online)Same as Fig. \ref{Cequ1} for $0.8<\lambda<0.9$.
\protect\label{Cequ}}
\end{figure}

\begin{figure}[htb]
\centering\includegraphics[width=10cm]{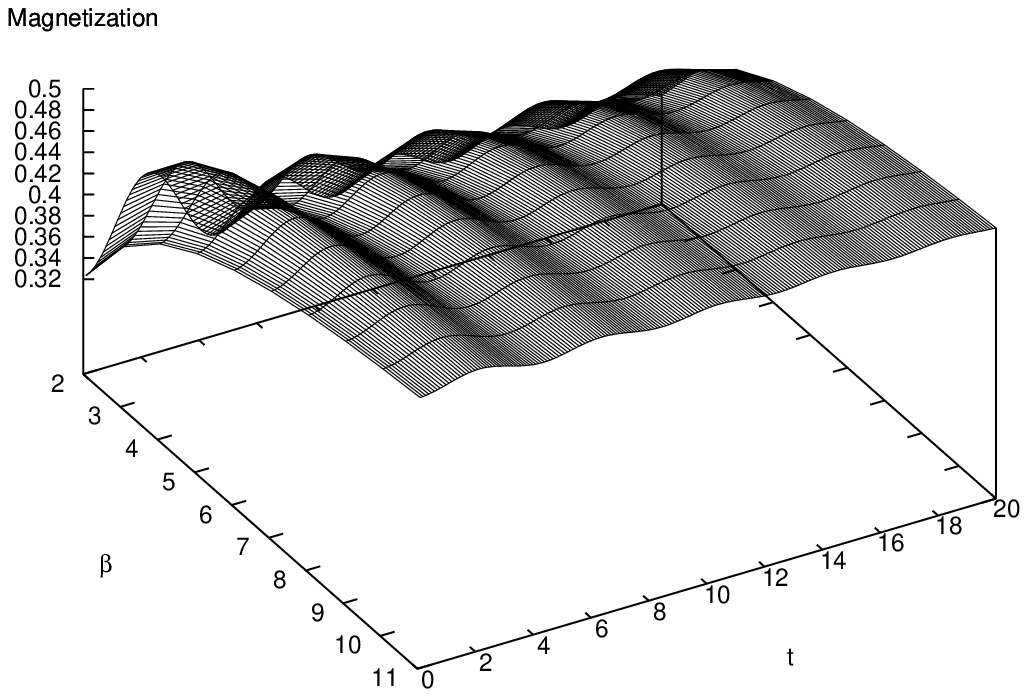}
\vspace{0cm}\caption{Site magnetization as function of the
temperature and time  with  $m-l=1$ and $\lambda=0.8$.
\protect\label{spin}}
\end{figure}

In  this section we present the solution of the N-sites Ising-XY
model with transverse field following the standard method
\cite{Lieb61,Baro70}. We proceed with the Hamiltonian
\be {\cal
H}=-\sum_{i=1}^{N}(\lambda[(1+\gamma)\sigma_i^x\sigma_{i+1}^x+(1-\gamma)\sigma_i^y\sigma_{i+1}^y]+\sigma_i^z)\,,\label{1}
\ee
where $\sigma^a_i$ are the Pauli operators, obeying the usual
commutation relations $[\sigma^a_i, \sigma^b_j] =
2i\epsilon^{abc}\delta^{ij}\sigma^c_i$ . The Zeeman energy in the
external magnetic field, as well as the Planck constant $\hbar$,
have been set to 1. We assume cyclic boundary conditions, i.e.,
the index $i$ in the sum (\ref{1}) runs over ${1\dots  N}$ with
$S_{N +1} = S_1$. This Hamiltonian can be diagonalized by means of
the Jordan-Wigner transformation \cite{Lieb61,Baro70} that maps
spins to one-dimensional spinless fermions with creation and
annihilation operators $c_i$ and $c_i^{\dag}$.
The Hamiltonian Eq.(1) in the fermionic operator
representation is represented by the quadratic form
\be
H=(\sum_{i,j}c_i^{\dag}A_{i,j}c_j+\frac{1}{2}[c_i^{\dag}B_{i,j}c_j^{\dag}+{\rm
H.c.}]) +N\,,\label{6}
\ee
where $A_{i,i}=-1$ and
$A_{i,i+1}=-\frac{1}{2}\lambda=A_{i+1,i}$,
$B_{i,i+1}=-\frac{1}{2}\lambda\gamma=-B_{i+1,i}$, and all other
$A_{i,j}$ and $B_{i,j}$ are zero. The quadratic Hamiltonian 
(\ref{6}) can be diagonalized by a linear Bogoliubov
transformation of the fermionic operators,
\be
\eta_k=\sum_i(g_{ki}c_i+h_{ki}c_i^{\dag})\,,\label{eta1}
\ee
\be
\eta_k^{\dag}=\sum_i(g_{ki}c_i^{\dag}+h_{ki}c_i)\label{eta2}\,,
\ee
where the $g_{ki}$ and $h_{ki}$ can be chosen to be real. After that it
 takes the diagonal form
\be
H=
\sum_k\Lambda_k\eta_k^{\dag}\eta_k-\frac{1}{2}\Lambda_k\,,
\ee
where
\be
\Lambda_k=\sqrt{(\gamma\lambda \sin k)^2+(1+\lambda \cos
k)^2}\,.
\ee

After the diagonalization of the Hamiltonian now we can proceed
with the evaluation of the thermal pairwise entanglement. The
pairwise entanglement, as its name indicates, measures how two
spins separated by a distance $r$ are entangled. This measure is
to be accomplished by evaluating the pairwise entanglement of the
two-site density matrix after tracing out all other spins in the
chain. We evaluate the entanglement of this states by using the
concurrence which is defined as~\cite{Benn96}
\be
{\cal
C}=\max\{\lambda_1-\lambda_2-\lambda_3-\lambda_4,0\}\,,
\label{eq:concur1}
\ee
where $\lambda$'s are the square roots of
the eigenvalues in decreasing order of the matrix
$\rho_{AB}(\sigma_y\otimes\sigma_y\rho_{AB}^{\star}
\sigma_y\otimes\sigma_y)$, where $\rho_{AB}^{\star}$ is the
corresponding complex conjugation in the computational basis
$\{|++\rangle, |+-\rangle, |-+\rangle,|--\rangle\}$. As usual, at
the thermal equilibrium the system is described by the canonical
ensemble density matrix
\be
\rho=\frac{e^{-\beta H}}{Z}\,,
\ee
where $Z={\rm Tr} e^{-\beta H}$ is the partition function of the
system. Thus  the  reduced two-site density matrix, after taking
into account the symmetries consideration, assumes the following
form
\ba
\rho_{ij}&=&{\rm Tr}_{k\neq i,j}\frac{e^{-\beta
H}}{Z}=\frac{1}{4}
( I\otimes I+\langle\sigma_z\rangle(\sigma_z^i\otimes I+I\otimes\sigma_z^j)\nonumber\\
&+&\sum_{k=1}^3\langle\sigma^{i}_k\sigma^{j}_k\rangle \sigma^{i}_k\otimes\sigma^{j}_k)\,.\label{dens}
\ea

The correlation functions that show up in the density matrix
Eq.(\ref{dens}) are well known \cite{Baro70} and for the nearest
neighbor case considering here these correlation
functions are giving by
\ba
&&\langle\sigma^{i}_x\sigma^{j}_x\rangle={\cal G}_{-1},\quad
\langle\sigma^{i}_y\sigma^{j}_y\rangle={\cal G}_{1},
\quad\langle\sigma^{i}_x\sigma^{j}_x\rangle=\langle\sigma_z\rangle^2-{\cal
G}_{1}{\cal G}_{-1},\nonumber\\&& \quad
\langle\sigma_z\rangle=-{\cal G}_{0}\,,
\ea
with
\ba
{\cal
G}_{mi}&=&\frac{1}{\pi}\int_0^{\pi}dk \cos[k(m-i)](1 + \lambda
\cos k)  \frac{\tanh (\Lambda_k\beta/2)}{\Lambda_k} \nonumber \\
&-& \frac{\lambda\gamma}{\pi}\int_0^{\pi}dk \sin[k(m-i)]\sin k
\frac{\tanh (\Lambda_k\beta/2)}{\Lambda_k} \,
\ea
Thus the
concurrence for the case of isotropic XY model, $\gamma=0$, reads
\be
{\cal C}=\max\{0,||{\cal G}_{1}|-\sqrt{\frac{1}{4}(1+{\cal
G}_{0}^2-{\cal G}_{1}^2)^2-{\cal G}_{0}^2}|\}\,,
\ee
and at $T=0$
we have ${\cal G}_0=1$ and
\[
{\cal G}_1=\left\{\begin{array}[c]{cc}0 &\quad { \lambda\leq 1}\\ \frac{2}{\pi}\sqrt{1-\lambda^{-2}}&\quad \lambda> 1\end{array} \right. \,.\label{rho}
\]
The two-site entanglement between the nearest-neighbors for the isotropic XY
model is shown in Fig.\ref{Cequ1}. One can see from this figure
that there is a region where the entanglement increases with the
temperature increase whereas for $\lambda \leq 1$ the entanglement
is zero at zero temperature and it remains zero until a critical
temperature where the system starts to be entangled.
Fig.\ref{Cequ} displays with more details a relevant region with
$0.8<\lambda<0.99$. It is clearly seen in this figure that there
is a strong enough entanglement in the region $2<\beta<20$,
however, outside this region no entanglement can be observed. This
entanglement transition should be understood as an effect of the
entanglement of the higher excited states. Note that the ground
state in this case is unentangled with all spins pointed in the
same direction. Similar observation has been made in
Refs.\onlinecite{Osbo02} and \onlinecite{31}. In the region where
the entanglement does not vanish we expect the quantum
fluctuations likely to dominate the behavior of the system and it
is the region where the quantum information processing should be
studied since it is well know that the entanglement is
the main resource for quantum information. Thus the study of the
dynamics of entanglement for this region is relevant. In the next
section we will study the response of the system after a
projective measurement in this region.

%%%%%%%%%%%%%%%%%%%%%%%%%%%%%%%%%%%%%%%%%%%%%%%%%%%%%%%%%%%%%%%%%%%%%%%%%%%%%%%%%%%%%%%%%%%%%%%%%%%%%%%%%%
\section{Spin decoherence after a projective measurement}
\label{sect:3}
\begin{figure}[htb]
\centering\includegraphics[width=10cm]{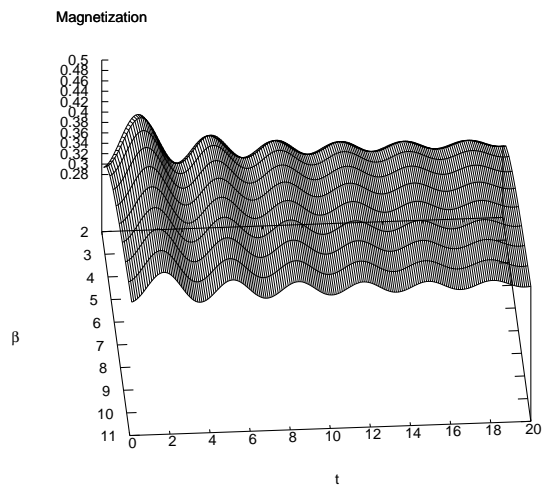}
\vspace{0cm}\caption{Site magnetization as function of the
temperature and time  with  $m-l=1$ and $\lambda=0.99$.
\protect\label{spin2}}
\end{figure}

\begin{figure}[htb]
\centering\includegraphics[width=10cm]{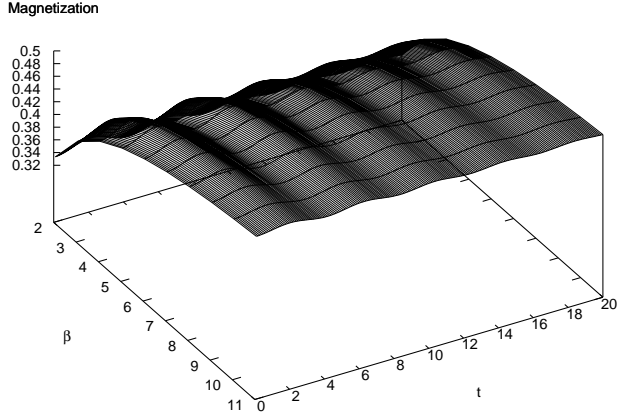}
\vspace{0cm}\caption{Site magnetization as function of the
temperature and time  with  $m-l=1$ and $\lambda=0.8$ for the case
without knowledge of the measurement outcome.
\protect\label{spinpm}}
\end{figure}

\begin{figure}[htb]
\centering\includegraphics[width=10cm]{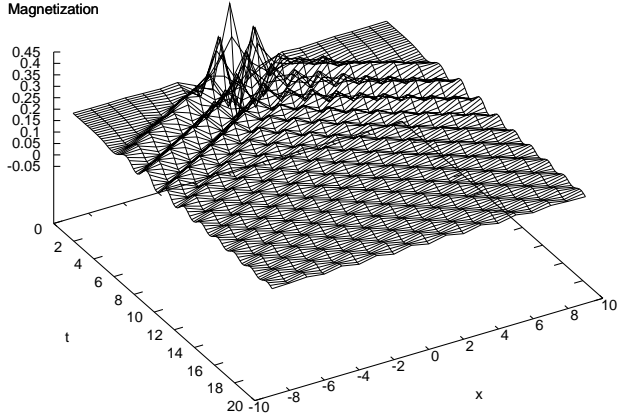}
\vspace{0cm}\caption{Site magnetization as function of the site
location $x=m-l$ and time  at fixed $\beta=10$ and $\lambda=2$.
\protect\label{spinT0.1L2}}
\end{figure}

As mentioned above we will study the consequences of the local
projective measurement and analyze the system behavior in terms of
the spin decoherence wave. We introduce the operators
\be
A_i=c_i^{\dag}+c_i\,\quad B_i=c_i^{\dag}-c_i\,.
\ee
After a
selective projective measurement with the projector
$P=\frac{\sigma_z^l+1}{2}=\frac{1-A_lB_l}{2}$ ($P=P^{\dag}$),
which means that the positive $z$ direction of the local spin $l$
is the measurement result (a general measurement will be
considered below), the mean value at time $t$ for an operator $A$
reads \cite{Hami04}
\be
\langle A(t)\rangle= \frac{{\rm Tr}\rho P
A(t)P}{{\rm Tr} P\rho P}\,,
\ee
where $A(t)=e^{iHt}A(0)e^{-iHt}$.
Since we are interested in the evaluation of the time-dependent
average value of the magnetization in the $z$ direction at site
$m$  we have $A(0)=\sigma_z/2=-\frac{A_mB_m}{2}$. In order to
evaluate $ \langle A(t)\rangle$ we write the operator $A_i$, and
$B_i$ in term of $\eta$ and $\eta{\dag}$ operators using the
inverse transformation of Eqs.(\ref{eta1}),(\ref{eta2}). Since the
Hamiltonian is diagonal being written in terms of $\eta$ operators
we have
\be
\eta^{\dag}_k(t)=\exp{(i\Lambda_kt)}\eta_k^{\dag}(0)\,.
\ee
Thus
after a straightforward little algebra we found the following
expression for $A(t)$

\begin{figure}[htb]
\centering\includegraphics[width=10cm]{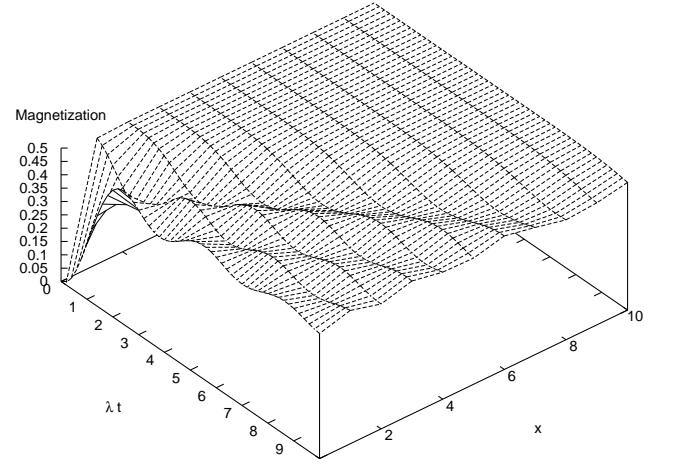} \vspace{0cm}
\caption{Site magnetization at distance $x$ from the measurement
point.\protect\label{magn}}
\end{figure}

\be
A(t)=\frac{-1}{2}\sum_{ii^{\prime}}\alpha_{ii^{\prime}}^mA_iA_{i^{\prime}}+\beta_{ii^{\prime}}^mA_iB_{i^{\prime}}+\gamma_{ii^{\prime}}^mB_iB_{i^{\prime}}\,,
\ee
where
\be
\alpha_{ii^{\prime}}^m=\phi_{mi}G_{mi^{\prime}};\,
\beta_{ii^{\prime}}^m=\phi_{mi}\psi_{mi^{\prime}}-G_{mi}G_{mi^{\prime}};\,
\gamma_{ii^{\prime}}^m=G_{mi}\psi_{mi^{\prime}}\,,
\ee
and in the
thermodynamic limit ($N\rightarrow \infty$) we have
\be
\phi_{\mu\nu}=\psi_{\mu\nu}=\frac{1}{\pi}\int_0^{\pi}dk\cos
k(\mu-\nu)\cos(\Lambda_kt)\,,
\ee
\ba
G_{\mu\nu}&=&\frac{i}{\pi}\int_0^{\pi}dk \cos[k(\mu-\nu)](1 + \lambda
\cos k) \frac{\sin (\Lambda_kt)}{\Lambda_k} \nonumber \\ &-&
\frac{i\lambda\gamma}{\pi}\int_0^{\pi}dk \sin[k(\mu-\nu)]\sin k
\frac{\sin (\Lambda_kt)}{\Lambda_k} \,.
\ea

We define the operator ${\cal
A}=\sum_{ii^{\prime}}\beta_{ii^{\prime}}^m A_iB_{i^{\prime}}$
which corresponds to the part of $A(t)$ with real coefficients.
Clearly we have $\langle{\cal A}\rangle_{\beta}=\langle
A\rangle_{\beta}$, therefore after some calculations we obtain
\ba
\langle A\rangle_{\beta} &=&\frac{1}{4({\cal
G}_{ll}+1)}\left.({\cal
G}_{mm}+\sum_{ii^{\prime}}\beta_{ii^{\prime}}^m[\langle
A_lB_lA_iB_{i^{\prime}}\rangle_{\beta}\nonumber\right.\\ &+&\left.
\langle A_iB_{i^{\prime}}A_lB_l\rangle_{\beta}-\langle A_lB_l
A_iB_{i^{\prime}}A_lB_l\rangle_{\beta}\right.])\,,
\ea
where
\be
\langle A_lB_lA_iB_{i^{\prime}}\rangle_{\beta}={\cal
G}_{ii^{\prime}}{\cal
G}_{ll}+\delta_{il}\delta_{i^{\prime}l}-{\cal G}_{li}{\cal
G}_{li^{\prime}}\,,
\ee
\be
\langle
A_iB_{i^{\prime}}A_lB_l\rangle_{\beta}={\cal G}_{ii^{\prime}}{\cal
G}_{ll}+\delta_{il}\delta_{i^{\prime}l}-{\cal G}_{il}{\cal
G}_{i^{\prime}l}\,,
\ee

\begin{widetext}
\ba
\langle A_lB_l A_iB_{i^{\prime}}A_lB_l\rangle_{\beta}=
-4\delta_{li^{\prime}}\delta_{il}{\cal G}_{ll}+\delta_{li}({\cal G}_{i^{\prime}l}+
{\cal G}_{li^{\prime}})+\delta_{li^{\prime}}({\cal G}_{il}+{\cal G}_{li})
+{\cal G}_{ll}{\cal G}_{i^{\prime}l}({\cal G}_{il}-{\cal G}_{li})
+{\cal G}_{ll}{\cal G}_{li^{\prime}}({\cal G}_{li}-{\cal G}_{il})-{\cal G}_{ii^{\prime}}\,.
\ea
\end{widetext}

Here we have used the Fermi distribution function
\be
\langle\eta_k\eta^{\dag}_{k^{\prime}}\rangle=\frac{\delta_{kk^{\prime}}}{e^{-\beta\Lambda_k}+1}\,.
\ee
For simplicity, in this paper we will further consider only
the isotropic XY Model, $\gamma=0$; the case $\gamma\neq 0$ will
be addressed in the future. Thus, the magnetization at the site
$m$ in the $z$ direction can be written as follow

\begin{widetext}
\ba
\langle{A}\rangle_{\beta}=\frac{1}{4({\cal G}_{ll}+1)}\left\{{\cal G}_{mm}+(2{\cal G}_{ll}+1)[2(\phi_{ml}^2-G_{ml}^2)+\alpha-\alpha^{\prime}]\right.+\left.4(G_{ml}\beta_{ml}^{\prime}-\phi_{ml}\beta_{ml})+2(\beta^{\prime 2}_{ml}-\beta_{ml}^2)\right\}\,,
\ea
\end{widetext}
where the expressions of
$\alpha,\,\beta_{ml},\,\alpha^{\prime},\,\beta_{ml}^{\prime}$ are
given in the Appendix.

The magnetizations at the neighboring site $m-l=1$ for the cases
$\lambda=0.8$ and $\lambda=0.99$ are shown in Figs.\ref{spin} and
\ref{spin2}, respectively, as functions of the inverse temperature
$\beta$ and of the time. The magnetization oscillate in time with
the frequency proportional to $\lambda$ which is a result of the
propagation of the decoherence wave after the local measurement
\cite{ourbec}. One can see that the amplitude of these
oscillations increases in the close vicinity of the quantum
critical point $\lambda=1$. Figs.\ref{spin} and \ref{spin2}
demonstrate also that the amplitude increases with the temperature
increase. It is connected probably with the thermal entanglement
in the system under consideration (see the previous section).

For a complete von Neumann measurement \cite{6} without knowledge
of the measurement outcome the mean value for the operator $A$ at
time $t$ is
\be
\langle A(t)\rangle= {{\rm Tr}\rho P A(t)P}+ {{\rm
Tr}\rho (1-P) A(t)(1-P)}\,,
\ee
Thus we have
\ba
\langle{A}\rangle_{\beta}&=&\frac{1}{4}\left\{{\cal G}_{mm}+4{\cal
G}_{ll}(\phi_{ml}^2-G_{ml}^2)+(\alpha-\alpha^{\prime})\right.\nonumber
\\&+&\left.4(G_{ml}\beta_{ml}^{\prime}-\phi_{ml}\beta_{ml})\right\}\,,
\ea

The magnetization at the neighboring site $m-l=1$ after the
measurement is plotted in Fig.\ref{spinpm} for $\lambda=0.8$ as a
function of $\beta$ and the time. One can clearly see from this
figure that there is a reduction of the amplitude of the
oscillation with respect to the selective measurement; this effect
is due to the mixing of the state as here we don't know the
results of the measurement outcome. Fig.\ref{spinT0.1L2} displays
the propagation of the decoherence wave, that is, the
magnetization distribution as a function of the distance from the
measured site $x$ and time $t$ at a fixed value of $\beta=10$ and $\lambda=2$.

\begin{figure}[htb]
\centering\includegraphics[width=10cm]{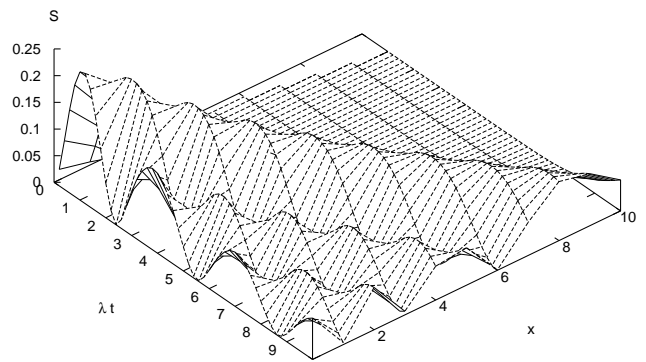}\vspace{0cm}
\caption{Single-site entanglement $S$ at distance $x$ from the
measurement point.\protect\label{sing}}
\end{figure}

\begin{figure}[htb]
\centering\includegraphics[width=10cm]{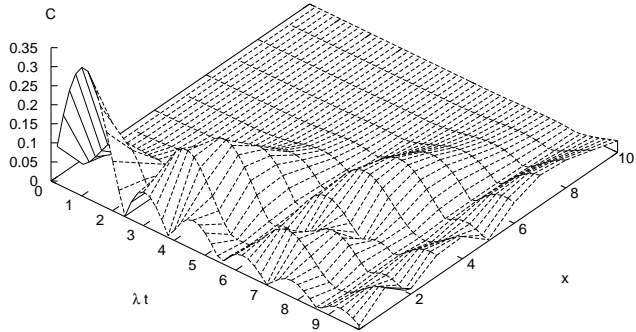}\vspace{0cm}
\caption{Two-site entanglement C between the site $m$ and the site
$i$ at zero temperature with $x=i-m$.\protect\label{twos}}
\end{figure}

\begin{figure}[htb]
\centering\includegraphics[width=10cm]{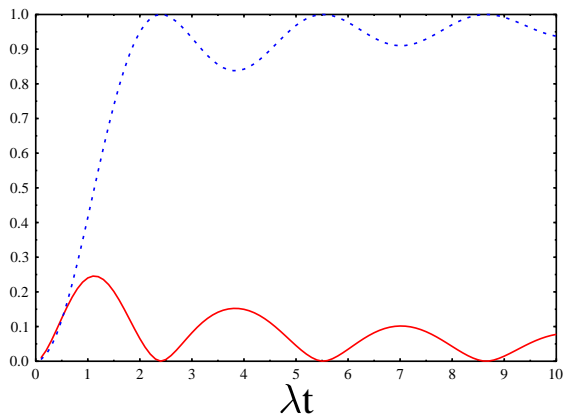} \vspace{0cm}
\caption{(color online) $\langle \sigma_z\rangle$ (dotted line) and single site
entanglement (solid line) are shown at the site
$m$.\protect\label{decoh}}
\end{figure}

\begin{figure}[htb]
\centering\includegraphics[width=10cm]{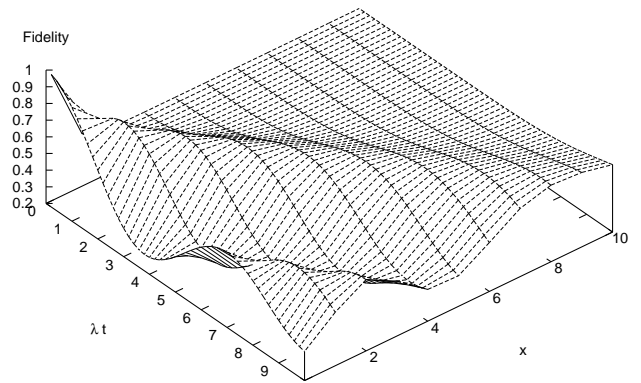} \vspace{0cm}
\caption{The fidelity of the quantum channel is shown at site
$i$.\protect\label{fide}}
\end{figure}

Now we consider the effect of the local measurement on the
pairwise entanglement in the system. This require the evaluation
of a correlations functions such as
$\langle\sigma_i^{\alpha}\sigma_j^{\beta}\rangle$. Here we will
present the results for an interesting particular case, namely,
for $\lambda<1$ and at zero temperature. Then, there is no
entanglement in the system before the measurement since all spin
are pointing in the same $z$-direction in the ground state.
Therefore a projective measurement of the $z$-component of the
magnetization will not provide us any nontrivial information and
will not generate entanglement. However, we will show that the
projective measurement of the $x$-component does create the
entanglement. After the projective measurement in the $x$
direction at site $m$ with positive outcome the wave function will
be
\be
|\Psi\rangle_{m}
 =\frac{1+c_m^{\dag}}{\sqrt 2}|vac\rangle\,.
\label{psi11}
\ee
At time $t$ we have
\be
c^{\dag}_m(t)=\sum_l(G_{ml}+\phi_{ml})c^{\dag}_l=\sum_lw_l(t)c^{\dag}_l\,,
\ee
where in the thermodynamics limit $w_l(t)=J_{m-l}(\lambda t)$
and $J_n(x)$ is the Bessel function of order $n$. Thus the
time-dependent wave function after the measurement will be \be
|\Psi_m(t)\rangle=\frac{1+\sum_l
w_l(t)c_l^{\dag}}{\sqrt{2}}|vac\rangle\,, \ee Thus, the
time-dependent two-spin density matrix can be written in the form
\[\rho_{ij}=\frac{1}{2}\left(\begin{array}[c]{cccc}0 & 0 & 0 & 0\\0 & w_i^2 & w_iw_j & w_i\\0 & w_iw_j & w_j^2 & w_j\\0 & w_i & w_j & 2-w_i^2-w_j^2\end{array}\right)  \,,\label{rho}\]
and the corresponding one-particle density matrix is
$\rho_{i}=\frac{1}{2}\left(\begin{array}[c]{cc}w_i^2 & w_i \\w_i&
2-w_i^2\end{array}\right)\,.$ The magnetization of the site $i$ is
therefore given by $\langle \sigma_z\rangle/2={\rm Tr }\rho
\sigma_z/2=(w_i^2-1)/2$. The single-site entropy,  $S(\rho_i)=-{\rm Tr}
\rho_i\log \rho_i$, which characterizes the entanglement of one
spin with the rest of the chain, can also be evaluated in this
case. The pairwise entanglement for the two-site density matrix
can be evaluated using Eq.(\ref{eq:concur1}). A straightforward
algebra lead to the following expression for the concurrence:
\be
{\cal C}=w_iw_j\,.
\ee
In Figs.\ref{magn}, \ref{sing}, \ref{twos}
we show the single-spin quantum entropy and the magnetization at
distance $x$ from the measurement point, as well as the pairwise
entanglement between site $m$ and $i$ as functions of time and of
$x=i-m$. We conclude from these figures that the single-site
entanglement and the pairwise entanglement propagate with the
velocity proportional to the interaction strength, like the spin
decoherence wave. In Fig.\ref{decoh} we display the site
magnetization together with the single-site entanglement which
demonstrates clearly that these two quantities oscillate
coherently. This confirms that the dynamics of the spin decoherence
reflects in some sense the dynamics of the entanglement in the
system.

The fidelity of the communication at site $m$ through the channel
is the probability that a channel output pass a test for being the
same as the input conducted by someone who knows what the input
was. It can be defined as \cite{Benn196}

\be F= {\langle}
\psi_m|\rho_i|\psi_m\rangle=\frac{1+w_i}{2}\,, \ee

where $|\psi_m\rangle=\frac{|0\rangle+|1\rangle}{\sqrt{2}}$ is the state of the site $m$
right after the measurement. In
fact, the spin chain acts as an amplitude damping quantum channel
where the initial state is transformed under the action of the
superoperator \$ to \cite{6}
\be
\rho\rightarrow \$(\rho)=M_0\rho
M_0^{\dag}+M_1\rho M_1^{\dag}\,,
\ee
with the Kraus operators such
$M_0=\left(\begin{array}[c]{cc}w_i & 0 \\0&
1\end{array}\right)\,,$ and $M_1=\left(\begin{array}[c]{cc}0 & 0
\\ \sqrt{1-\omega_i^2}&
0\end{array}\right)\,$ where as usual $M_1$ describe the quantum
jump and $M_0$ represent no quantum jump.  The fidelity of the channel is shown in Fig.\ref{fide}.
One can see clearly from this figure that the channel can be
efficiently used to transmit the quantum information. The fidelity
has a maximum value for $x=i-m\sim \lambda t$. This means that the
quantum state is transported with the velocity proportional to the
interaction strength $ \lambda$ similar to the decoherence wave.
After a time $t=x/\lambda$  the state can be recovered with
maximum fidelity at a distance $x$ from the initial site $m$. 

\section{Conclusions}
\label{sect:4}

In this paper, we have evaluated the equilibrium pairwise entanglement at
finite temperatures in the isotropic one-dimensional Ising-XY model
with transverse magnetic field. Our findings indicate that the behavior
of entanglement with respect to temperature, at least for moderate
values of temperature, is quite complex. In particular, we have found that
for some ranges of temperature, entanglement in the system can grow with
increasing temperature, which results from the entanglement of
the excited states. We have studied the dynamical response of
the system in a relevant region  for quantum information processing after
a projective measurement on one local spin  which leads to the
appearance of the ``decoherence wave'' that propagate with 
velocity proportional to the coupling constant $\lambda$, similar
to the case of the Bose-Einstein condensate studied earlier \cite{ourbec}. 
One motivation behind our study is to know what happens with the 
quantum computer after the measurement. We have investigated
for specific case ($T=0$ and $\lambda<1$) the dynamics of 
the  entanglement and the spin decoherence wave and we have found that
those quantities propagates coherently through the chain with the
same velocity that is proportional to $\lambda$. The fidelity of the channel 
has been shown to be  represented as amplitude damping channel. Finally, a generalization to
$\gamma\neq 0$ and study of the entanglement dynamics in such system are desirable.

\section*{Appendix}

Using the standard properties of the Fourier transformation, one
has

\begin{widetext}
\ba
\sum_lG_{il}\bar{G}_{i^{\prime}l}=\sum_l\int_{-\pi}^{\pi}\frac{dk}{2\pi}G(k)e^{ik(i-l)}\int_{-\pi}^{\pi}\frac{dk^{\prime}}{2\pi}\bar{G}(k^{\prime})e^{ik^{\prime}(i^{\prime}-l)} =\int_{-\pi}^{\pi}\frac{dk}{2\pi}G(k)\bar{G}(-k)e^{ik(i-i^{\prime})}
\ea

Putting $\gamma=0$ we find

\ba
 \alpha&=&\sum_{ii^{\prime}}\phi_{mi}\phi_{mi^{\prime}}{\cal G}_{ii^{\prime}}=\frac{1}{\pi}\int_0^{\pi}dk (1 + \lambda \cos k) \frac{\tanh (\Lambda_k\beta/2)}{\Lambda_k}\cos^2(\Lambda_kt) \nonumber \\
\ea

\ba
\beta_{ml}&=& \sum_i\phi_{mi}{\cal G}_{li}=\frac{1}{\pi}\int_0^{\pi}dk \cos[k(m-l)](1 + \lambda \cos k) \frac{\tanh (\Lambda_k\beta/2)}{\Lambda_k}\cos(\Lambda_kt) \nonumber \\
\ea

\ba
 \alpha^{\prime}&=&\sum_{ii^{\prime}}G_{mi}G_{mi^{\prime}}{\cal G}_{ii^{\prime}}=\frac{-1}{\pi}\int_0^{\pi}dk (1 + \lambda \cos k)^3 \frac{\tanh (\Lambda_k\beta/2)}{\Lambda_k^3}\sin^2(\Lambda_kt) \nonumber \\
\ea

\ba \beta_{ml}^{\prime}&=& \sum_iG_{mi}{\cal G}_{li}= \frac{i}{\pi}\int_0^{\pi}dk \cos[k(m-l)](1 + \lambda \cos k)^2 \frac{\tanh (\Lambda_k\beta/2)}{\Lambda_k^2}\sin(\Lambda_kt) \nonumber \\
\ea
\end{widetext}

%%%%%%%%%%%%%%%%%%%%%%%%%%%%%%%%%%%%%%%%%%%%%%%%%%%%%%%%%%%%%%%%%%%%%%%%%%%%%%%%%%%%%%%%%%%%%%%
\section*{Acknowledgments}

This work was performed as part of the research program of the
\textsl{Stichting voor Fundamenteel Onderzoek der Materie (FOM)}
with financial support from the \textsl{Nederlandse Organisatie
voor Wetenschappelijk Onderzoek }.


\begin{thebibliography}{99}

\bibitem{1} J. S. Bell, Physics {\bf 1}, 195 (1964);
Rev. Mod. Phys. {\bf 38}, 447 (1966).

\bibitem{2} A. Aspect, Nature (London) {\bf 398}, 189 (1999).

\bibitem{3} M. Lamehi-Rachti and W. Mittig, Phys. Rev. D {\bf 14}, 2543 (1976).

\bibitem{4}C. Polachic, C. Rangacharyulu, A.M. van den Berg, S. Hamieh, M.N. Harakeh, M. Hunyadi,
M.A. de Huu, H.J. W\"ortche, J. Heyse, C. Baumer., D. Frekers, S.Rakers, J.A. Brooke and P. Busch, Phys. Lett. A {\bf 323}, 176 (2004).
\bibitem{44}S. Hamieh, H.J. W\"ortche, C. Baumer, A.M. van den Berg, D. Frekers, M.N. Harakeh,
 J. Heyse, M. Hunyadi, M.A. de Huu, C. Polachic and C. Rangacharyulu, J. Phys. G {\bf 30}, 481 (2004).

\bibitem{5}S. Sachdev, {\it Quantum Phase Transitions} (Cambridge
University Press, Cambridge, 1999).

\bibitem{Osbo02} T. Osborne and M. Nielsen,
Phys. Rev. A {\bf 66}, 032110 (2002).

\bibitem{19} A. Osterloh, L. Amico, G. Falci, and R. Fazio,
Nature (London) {\bf 416}, 608 (2002).

\bibitem{Hami05} S. Hamieh and A. Tawfik, Acta Phys.
Polon. B {\bf 36}, 801 (2005).

\bibitem{12} M. A. Nielsen, Ph.D. thesis, University of New
Mexico, 1998; quant-ph/0011036.

\bibitem{14} P. Zanardi and X. Wang, J. Phys. A {\bf 35}, 7947 (2002).

\bibitem{24} X. Wang, H. Fu, and A. I. Solomon, J. Phys. A {\bf 34},
11307 (2001).

\bibitem{25} W. K. Wootters, Contemporary Mathematics {\bf 305}, 299 (2002).

\bibitem{26} M. C. Arnesen, S. Bose, and V. Vedral,
Phys. Rev. Lett. {\bf 87}, 017901 (2001).

\bibitem{27} D. A. Meyer and N. R. Wallach, J. Math. Phys. {\bf 43}, 4273 (2002).

\bibitem{28} D. Gunlycke, S. Bose, V. M. Kendon, and V. Vedral,
Phys. Rev. A {\bf 64}, 042302 (2001).

\bibitem{29} X. Wang, Phys. Rev. A {\bf 64}, 012313 (2001).

\bibitem{30} H. Fu, A. I. Solomon, and X. Wang, J. Phys. A {\bf 35}, 4293 (2002).

\bibitem{31} X. Wang, Phys. Lett. A {\bf 281}, 101 (2001).

\bibitem{32} X. Wang and P. Zanardi, Phys. Lett. A {\bf 301}, 1 (2002).

\bibitem{Amic04}  L. Amico, A. Osterloh, F. Plastina, R. Fazio, and G. M. Palma,
Phys. Rev. A {\bf 69}, 022304 (2004).

\bibitem{Subr05} V. Subrahmanyam and A. Lakshminarayan, quant-ph/0409048;
V. Subrahmanyam, Phys. Rev. A {\bf 69}, 022311 (2004); V.
Subrahmanyam, Phys. Rev. A {\bf 69}, 034304 (2004).

\bibitem {ghos1} S. Ghosh, T. F. Rosenbaum, G. Aeppli, and S. N.
Coppersmith, Nature (London) {\bf 425}, 48 (2003).

\bibitem{gu} S. Gu, S. Deng, Y. Li, and H. Lin,
Phys. Rev. Lett. {\bf 93}, 086402 (2004).

\bibitem{Anfo} A. Anfossi, C. Boschi, A. Montorsi,
and F. Ortolani, cond-mat/0503600.

\bibitem{ourkondo} M. I. Katsnelson, V. V. Dobrovitski, H. A. De Raedt,
and B. N. Harmon, Phys. Lett. A {\bf 318}, 445 (2003).

\bibitem{6}M. Nielsen and I. Chuang, {\it Quantum Computation
and Quantum Information} (Cambridge University Press, Cambridge,
2000).

\bibitem{7} S. Bose, Phys. Rev. Lett. {\bf 91},  207901 (2003).

\bibitem{Benn96} C.  Bennett, D.  DiVincenzo, J. Smolin, and  W. Wootters,
Phys. Rev. A {\bf 54}, 3824 (1996); S. Hill and W. Wootters, Phys.
Rev. Lett. {\bf 78}, 5022 (1997); W. Wootters, Phys. Rev. Lett.
{\bf 80}, 2245 (1998).


\bibitem{ourbec} M. I. Katsnelson, V. V. Dobrovitski, and B. N. Harmon,
Phys. Rev. A {\bf 62}, 022118 (2000).

\bibitem{ourneel} M. I. Katsnelson, V. V. Dobrovitski, and B. N. Harmon,
Phys. Rev. B {\bf 63}, 212404 (2001).

\bibitem{Lieb61} E. Lieb, T. Schultz, and D. Mattis, Ann. Phys. (N.Y.) {\bf 60}, 407 (1961).

\bibitem{Baro70} E. Barouch and B. McCoy, Phys. Rev. A {\bf 2}, 1075 (1970);
E. Barouch and B. McCoy, Phys. Rev. A {\bf 3}, 786 (1971).

\bibitem{Hami04} S. Hamieh, R. Kobes, and H. Zaraket, Phys. Rev. A {\bf 70}, 052325 (2004).

\bibitem{Benn196} C. Bennett, H. Bernstein, S. Popescu, and B.
Schumacher, Phys. Rev. A {\bf 53}, 2046 (1996).

\end{thebibliography}
\end{document}